\documentclass[twocolumn]{aastex62}
\usepackage{hyperref}
\usepackage{cleveref}

\graphicspath{{./}{figures/}}

\received{January 1, 2018}
\revised{January 7, 2018}
\accepted{\today}
\submitjournal{ApJL}

\shorttitle{Extended coronal emission in Circinus}
\shortauthors{Rod\'{\i}guez-Ardila \& Fonseca-Faria}

\begin{document}

\title{A 700-pc extended coronal gas emission in the Circinus galaxy}

\correspondingauthor{Alberto Rodr\'{\i}guez-Ardila}
\email{aardila@lna.br}

\author[0000-0002-0786-7307]{Alberto Rodr\'{\i}guez-Ardila}
\affil{Laborat\'orio Nacional de Astrof\'{\i}sica. 
Rua dos Estados Unidos, 154, CEP 37504-364 \\
Itajub\'a, MG, Brazil}
\affil{Instituto Nacional de Pesquisas Espaciais. Divis\~ao de Astrof\'isica. Avenida dos Astronautas 1758.\\ S\~ao Jos\'e dos Campos, 12227-010, SP, Brazil}
\author{Marcos A. Fonseca-Faria}
\affil{Instituto Nacional de Pesquisas Espaciais. Divis\~ao de Astrof\'isica. Avenida dos Astronautas 1758.\\ S\~ao Jos\'e dos Campos, 12227-010, SP, 
Brazil}
\nocollaboration



\begin{abstract}

We report the first characterization of an extended outflow of high ionized gas in the Circinus Galaxy by means of the coronal line [\ion{Fe}{7}]~$\lambda$6087~\AA. This emission is located within the ionization cone already detected in the [\ion{O}{3}]~$\lambda$5007~\AA\ line and is found to extend up to a distance of $\sim$700~pc from the AGN. The gas distribution appears clumpy, with several knots of emission. Its kinematics is complex, with split profiles and line centroids shifted from the systemic velocity. The physical conditions of the gas show that the extended coronal emission is likely the remnants of shells inflated by the passage of a radio-jet. This scenario is supported by extended X-ray emission, which is spatially coincident with the morphology and extension of the [\ion{Fe}{7}]~$\lambda$6087~\AA\ gas in the NW side of the galaxy. The extension of the coronal gas in the Circinus galaxy is unique among active galaxies and demonstrates the usefulness of coronal lines for tracing the shock ionization component in these objects.

\end{abstract}



\section{Introduction} \label{sec:intro}

The relevance of the kinetic channel as a major way of releasing nuclear energy to the ISM in low-luminosity active galactic nuclei (AGN) has been so far  underestimated \citep{mh13,wm18}. The lack of convincing observational evidence of strong feedback in such a class  of objects has been, up to very recently, scarce.  
 
Recent works, however, reveal that low-power radio jets can play a major role in driving fast, multi-phase, galaxy-scale outflows even in radio-quiet AGN \citep{rodri+17,may+18,harvis+19}. Mass outflow rates, of up to 8 M$\odot$ yr$^{-1}$, similar to that found in powerful radio-loud AGN have been derived. Overall, they all point towards the relevance of low Eddington sources in delivery kinetic energy to the ISM. 
  
Traditionally, the identification of outflows in the warm, ionized phase of AGN has been done by means of the [\ion{O}{3}]~$\lambda$5007~\AA\ line \citep{greene+11}. In combination with integral field spectroscopy, critical insights about the geometry, structure, extension and physical conditions of the outflowing gas can be derived. This technique has allowed the discovery of kiloparsec-scale outflows in AGN samples \citep{hump+10,harrison+14,karouzos+16}. 

However, because [\ion{O}{3}] is also emitted in the galaxy disk and star-forming regions, isolating the contribution due to outflows is not  straightforward. In this respect, \citet{rodri+06} showed that high-ionization lines such as [\ion{Fe}{7}]~$\lambda$6087 in the optical or [\ion{Si}{6}]~1.963$\mu$m in the near-infrared (NIR) are excellent tracers of the ionized component of the outflows. Both lines are usually the brightest coronal lines (CLs) in their respective wavelength regimes. The energy required for their production ($\chi \geq 100$~eV, where $\chi$ is the ionization potential required to produce the ion) rules out stellar or galactic origin. The case of ESO~428-G\,14 \citep{may+18} is emblematic. These authors employed the [\ion{Si}{6}] to trace the jet-driven mechanical energy and the corresponding mass outflow deposited by the jet in the central 170 parsecs of that object.  

At the adopted distance of 4.2 Mpc  (1" $\sim$ 20.4 pc), Circinus is the closest Seyfert~2 galaxy to us.  Because of its proximity, angular resolution on scales of a few tens of parsecs can be reached even at seeing-limited conditions. It makes Circinus an excellent laboratory to study the physics of the ionized gas at a large range of distances. This includes the close examination of one of the most prominent characteristics of this object: an spectacular one-sided ionization cone, seen mostly in the NW side of the galaxy using the [\ion{O}{3}]~$\lambda5007$ line that extends for at least 30$\arcsec$ \citep{elmo+98}. 

In addition, Circinus is widely known for its prominent CL spectrum \citep{moor+96}.
\citet{prieto+05} determined the size and morphology of the coronal line region (CLR) for the first time in Circinus by means of adaptic (AO) imaging observations. Using [\ion{Si}{7}]~2.48$\mu$m as a tracer of the CLR, they found that the region emitting that line  extends from the nucleus up to 30~pc.  Later, \citet{muller+06} and \citet{muller+11} using $K-$band SINFONI/VLT AO integral field unit (IFU) spectroscopy revealed extremely strong coronal emission lines of [\ion{Si}{6}] and [\ion{Ca}{8}] as well as [\ion{Al}{9}].   In all three coronal lines, in addition to an unresolved, narrow component, they found a broad (FWHM $>$ 300~km~s$^{-1}$) blue-shifted component, which is spatially extended. It was argued that the narrow component arises in clouds physically close to the AGN while the blue wing originates from cloudlets that have been eroded from the main clouds and are accelerated outward. Due to the small field-of-view (FoV) of the detector (3" $\times$ 3"), it was not possible to determine if  extended CL emission out of the nuclear region was present in Circinus.

Here, we report optical observations of the Circinus Galaxy focusing on two main aspects: (i) the determination of the full extension of the  coronal gas and its relationship with galactic feedback; (ii) the role of a jet in shaping the morphology of the high-ionization gas in this object.

\section{Observations and data analysis} \label{sec:obs}

Data for Circinus, observed using MUSE/VLT  was retrieved from the ESO science portal. They were obtained under an equivalent airmass of 1.3, average seeing of 0.78$\arcsec$ and exposure time of 1844~s. 
The MUSE IFU cube consists of $\sim 300 \times  300$ spaxels, for a total of over 90000 spectra with a spatial sampling of 0.2$\arcsec \times 0.2\arcsec$/spaxel. The FoV of 1’$\times$1’ covers the central part of the galaxy, spanning an area of $\sim1.2\times 1.2$ kpc$^2$. The spectral resolution varies from 1750 at 4650 \AA\ to 3750 at 9300 \AA. 
The IFU cube, as retrieved from the science portal, is fully reduced, including calibration in flux (in absolute units) and wavelength.
   
The MUSE datacube was analyzed making use of a set of custom python scripts developed by us as well as software publicly available in the literature. First, we rebinned the cube to a spatial sampling of 0.6$\arcsec \times 0.6\arcsec$, reducing the total number of spaxels to $\sim$10000. This procedure was employed to increase the signal-to-noise ratio (SNR) of weak emission lines within the ionization cone. As the angular extension of the features we plan to study are of several arcsecs in size, the new spaxel scale does not affect the results reported here. 

We then remove the stellar continuum across the whole spectral range of MUSE (4700 $-$ 9100~\AA) using {\sc{starlight}} \citep{cid+05} and the \citet{bc03} stellar libraries. 

After this procedure, we measured the integrated emission line fluxes of H$\alpha$ and H$\beta$ at every spaxel to determine the extinction (Galactic and intrinsic) affecting the gas. This was done assuming an intrinsic line ratio H$\alpha$/H$\beta =$ 3.1 and the \citet{ccm87} extinction law. 
Our results point out to a Galactic extinction, $A_{\rm V}$, of 2.1 mag, in agreement to that reported by \citet{mingozzi+19}. The internal extinction varies from nearly zero to 2 mag. In the inner region of the ionization cone, $A_{\rm V}$  is rather homogeneous and low, with values between 0.1-0.3 mag.  Integrated line fluxes measured at each spaxel were corrected by the extinction measured accordingly.

Finally, for each spaxel, we fit the [\ion{Fe}{7}]~$\lambda$6087~\AA\ line with single or a combination of Gaussian functions. Typically, one or two components  were necessary to reproduce the observed profile. This procedure allowed us to measure the flux, velocity and velocity dispersion of the line across the whole MUSE field-of-view. 

\section{Morphology and kinematics of the coronal gas} \label{sec:outflow}

\begin{figure*}
\gridline{\fig{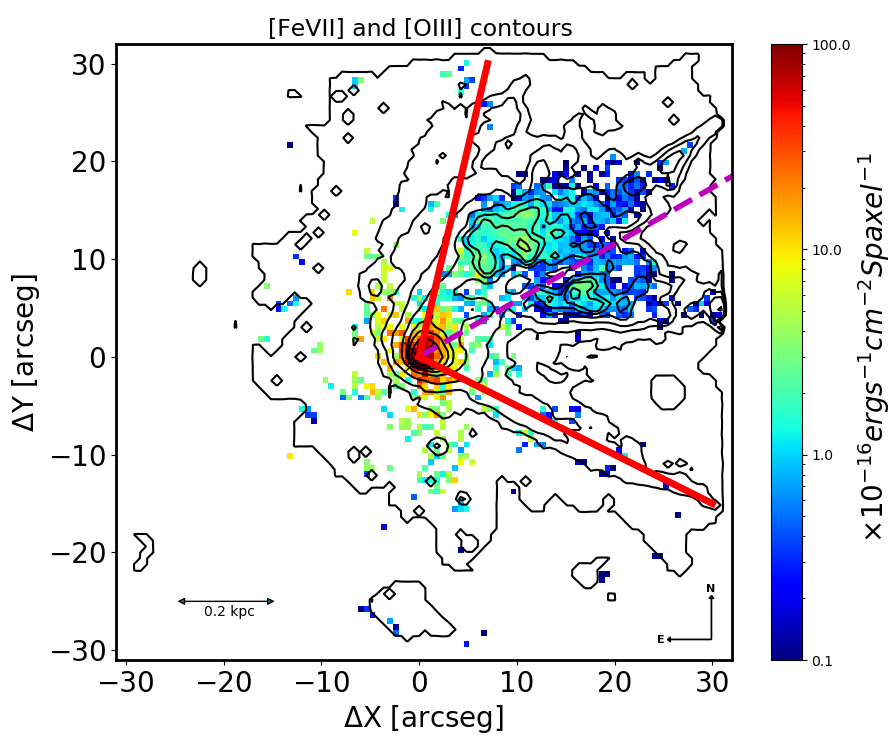}{0.32\textwidth}{(a)}
          \fig{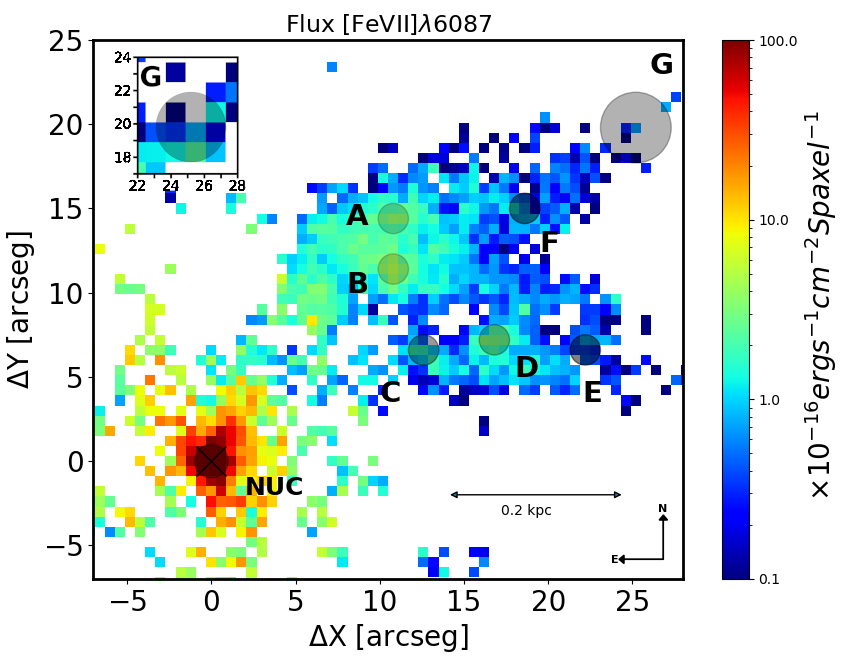}{0.34\textwidth}{(b)}
          \fig{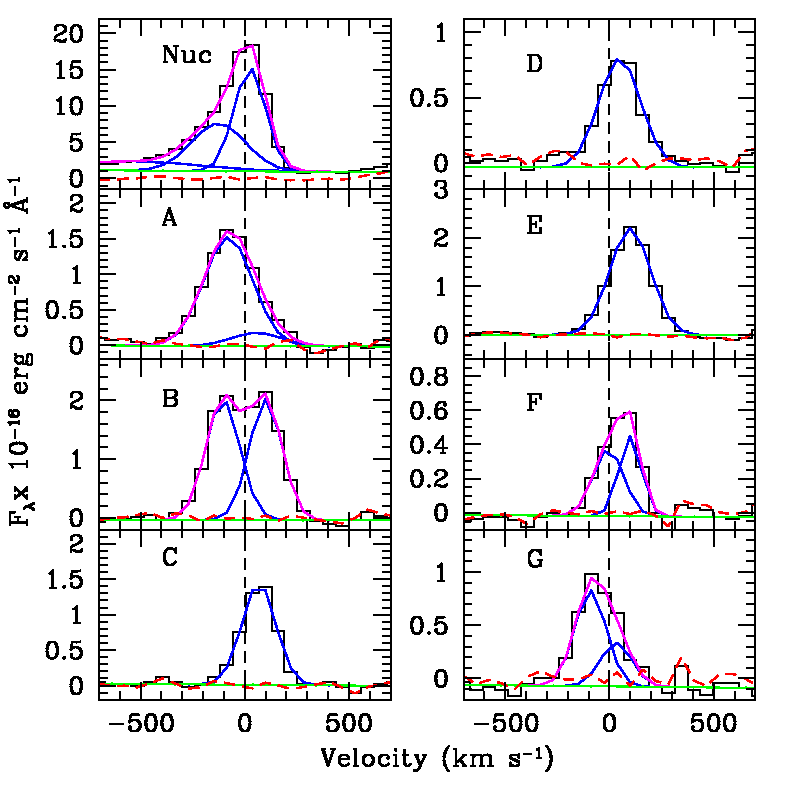}{0.32\textwidth}{(c)}
          }
\caption{(a): emission map of the [\ion{Fe}{7}]~$\lambda$6087~\AA, overlaid to [\ion{O}{3}]~$\lambda$5007~\AA\ contours. The red lines mark the edges of the ionization cone \citep{mingozzi+19}. The dashed magenta line indicates the PA=295$\pm5^{\rm o}$ of the radio continuum reported by \citet{elmo+98}. (b): zoom to NW ionization cone, emphasizing the extended [\ion{Fe}{7}] emission. The  letters A to G indicate regions where the line profiles shown in panel (c) were extracted. The AGN position is labeled ``Nuc". The inset in the upper left corner shows region G, with the spaxels binned to a size of 1.2"$\times$1.2" to increase the S/N to $>$3. (c): selected [\ion{Fe}{7}]~$\lambda$6087~\AA\ line profiles, in velocity space, detected in the extended HIG. The dashed vertical line marks the systemic velocity of the galaxy. On each panel, the observed profile (black histogram) was fit with Gaussian functions (blue lines). The the total fit is in magenta, the continuum level is in green and the residual after subtraction of the total fit to the observed data is in dashed red.
\label{fig:fe7map}}
\end{figure*}

A careful inspection of each spaxel in the datacube of the Circinus Galaxy reveals a very rich emission line spectrum, with prominent [\ion{O}{3}]~$\lambda$5007~\AA\ and H$\alpha$ lines across most of the FoV covered by MUSE. 
\citet{mingozzi+19} recently presented the gas distribution for these lines using the same dataset as here. It can be seen, for instance, the H$\alpha$ disc component flux map, overlaid with the [\ion{O}{3}]~$\lambda$5007~\AA\ outflow component as well as the ISM properties in the Circinus galaxy (see their figures 2 and 3, respectively).
In this letter, we concentrate on the 2D properties of the extended coronal gas of that AGN. \citet{oliva+99}, using optical spectroscopy, had already reported extended [\ion{Fe}{7}] emission up to 22$\arcsec$ from the center at a PA=318$^{\rm o}$. \citet{mingozzi+19} reported the presence of extended [\ion{Fe}{7}]~$\lambda6087$ associated to the ionization cone, but no information about its full extension, morphology and gas physical properties were presented.

Figure~\ref{fig:fe7map} shows, for the first time in the literature, the most complete [\ion{Fe}{7}]~$\lambda$6087~\AA\ emission line map from the MUSE datacube. White areas correspond to masked regions with SNR $<$3. It can be seen that the central region of the NW ionization cone is filled with a conspicuous, extended coronal emission. 
The size of the highly-ionized region occupies at least a projected area of 400 $\times$ 300~pc$^2$.  The gas distribution appears clumpy, arranged in several condensations, most of them contained within two main regions of emission, the latter two separated by a strip of very little or no [\ion{Fe}{7}] and coincident with the PA of the  radio-continuum (295$^{\rm o} \pm5^{\rm o}$, \citealt{elmo+98}). In the following, we will refer to this radio-continuum emission as the radio jet. The region to the North (hereafter NR), is elongated in the SE-NW direction, measuring approximately 200~pc  SE-NW and 135~pc NE-SW. It contains a bright spot (which \citealt{oliva+99} named ``knot C"), at $\sim 2\arcsec$ off-centre of NR, towards the bottom edge of the cloud. Here, it is identified by the letter B (see central panel of Figure~\ref{fig:fe7map}). 

The region to the South (hereafter SR) is composed of three smaller blobs of emission, (identified by the letters C to E in the central panel of Figure~\ref{fig:fe7map}) arranged nearly in the E-W direction.

In addition to the NR and SR, clouds of smaller size emitting [\ion{Fe}{7}]  ($< 20$~pc in radius) are also visible in the image. Most of them are located  outwards of the far edge of the NR. The farthest one is detected up to $\sim 34\arcsec$ from the nucleus ($\sim$700~pc from the AGN). This spot is labeled G in the central panel of Figure~\ref{fig:fe7map}. In order to clearly detect [\ion{Fe}{7}] at a S/N $>$3 in G, we further re-rebinned the spaxels to 1.2"$\times$1.2" in size and integrated the flux within a circular aperture of 4$\arcsec$ in radius. In the remainder of this manuscript, we will named the region enclosed by points A to G in Figure~\ref{fig:fe7map} as the extended emission. We do recognize the presence of extended, circumnuclear emission within a radius of $\sim 5\arcsec$ centred at the nucleus, but that is not the main focus of this work.

From the MUSE data, the integrated [\ion{Fe}{7}] flux in the circumnuclear region is 49.4$\times10^{-14}$ erg\,s$^{-1}$\,cm$^{-2}$. The extended CL emission, integrated across all the NW region (see panel b, Figure~\ref{fig:fe7map}), amounts to 6.5$\times10^{-14}$ erg\,s$^{-2}$\,cm$^{-2}$. These values are already corrected for extinction. Thus, the extended emission represents more than 10\% of the nuclear value.

\begin{figure*}
    \centering
    \includegraphics[width=\textwidth]{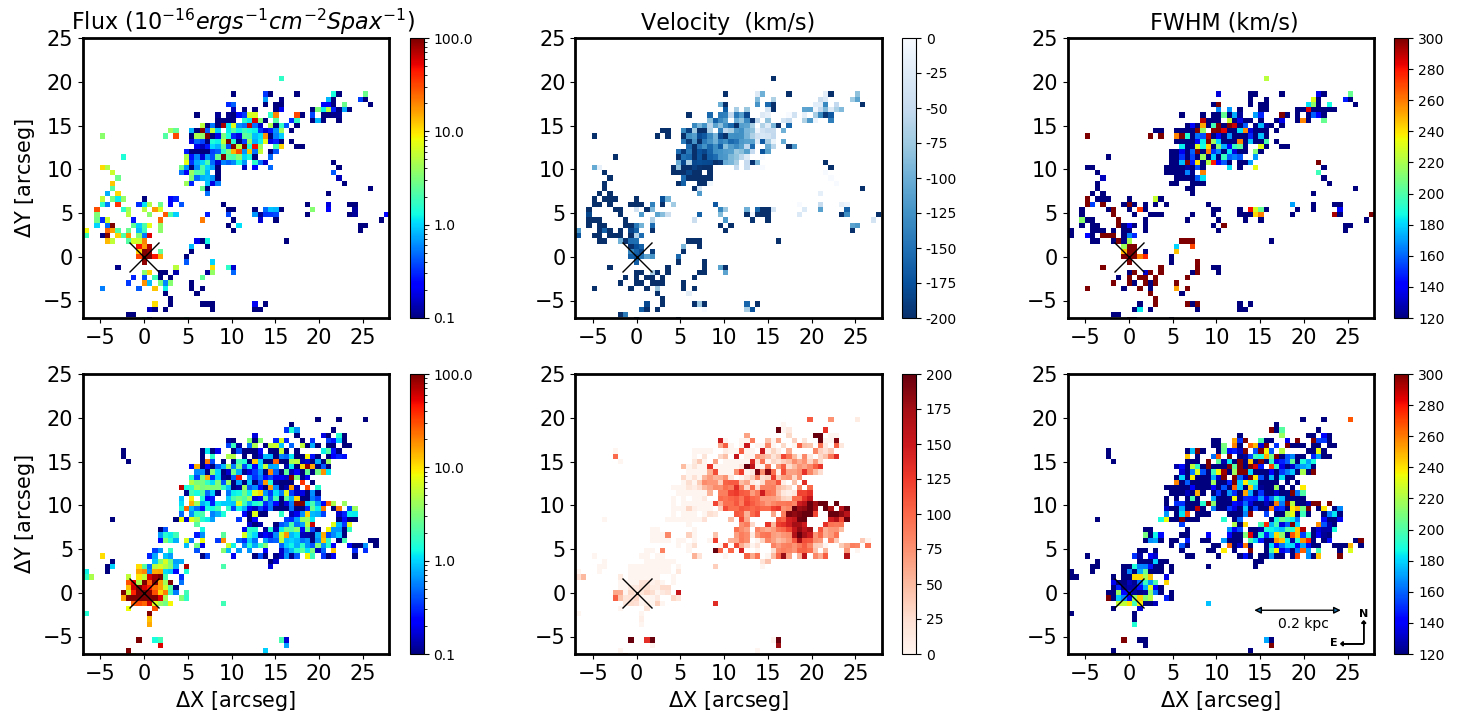}
    \caption{Flux distribution (left panel); velocity map (center panel) and FWHM of the line (right panel) for the blue and red component (upper and lower panels, respectively) of the [\ion{Fe}{7}]~6087~\AA\ line. The cross marks the position of the AGN.}
    \label{fig:kinematics}
\end{figure*}

The [\ion{Fe}{7}]~6087~\AA\ emission line profiles display a complex structure, with splitted lines at most spaxels in the NR. The relative intensity between the peaks and the width of the individual components vary appreciably. This is shown in panel (c) of Figure~\ref{fig:fe7map}, where  [\ion{Fe}{7}] emission line profiles extracted at different positions of the extended emission are plotted. Gaussian fitting carried out on these lines evidences the necessity of two components at positions A, B, F and G in order to adequately reproduce the observed profiles. The line profiles at positions C, D and E are mostly redshifted, with just the red peak visible Co-spatial to the [\ion{Fe}{7}] emission, lines of \ion{H}{1}, [\ion{O}{3}], [\ion{O}{2}], [\ion{N}{2}], [\ion{O}{1}], [\ion{S}{2}] and [\ion{S}{3}] are also observed in the corresponding integrated spectra (not shown here).

Figure~\ref{fig:kinematics} shows the integrated flux (left panel), velocity (central panel) and FWHM (right panel) for the blue component (upper row) and red component (bottom row) of the [\ion{Fe}{7}]~6087~\AA\ line.  All FWHM values were previously corrected in quadrature for the instrumental resolution. 
The velocity pattern in the extended emission observed in Figure~\ref{fig:kinematics} rules out that the coronal gas is rotating with the galaxy disk. This result is in agreement with previous works on the gas kinematics within the ionization cone of Circinus, pointing out that most of the material in that structure is out of the galaxy plane \citep{elmo+98}.

Figure~\ref{fig:kinematics} also shows that the FWHM of both blue and red components are in the interval  $\sim$160-200~km\,s$^{-1}$. At some spaxels, though, it jumps to $\sim$260~km\,s$^{-1}$ or higher values. 
At 80\% of the line width, the velocity reaches $\sim$350~km~s$^{-1}$ (see panel (c) in Figure~\ref{fig:fe7map}). Moreover, the relative separation between the peaks varies along the extended emission, from 130~km~s$^{-1}$ at regions F and G to nearly 250~km~s$^{-1}$ at A and B.

It is also interesting to see that the bulk of the coronal gas shows evidence of expanding shells. For the blue component, the shift of line decreases from $\simeq$ -200~km\,s$^{-1}$ in the region that is closest to the central source, to the systemic velocity at the outermost portion of the cloud. Thus, if the gas is travelling towards the observer, we see preferentially the portion closest to us. For the red peak, receding velocities of up to $\sim$200~km\,s$^{-1}$ are detected. These clouds are located preferentially in the regions along the PA of the radio-jet.

\section{Discussion} \label{sec:final}

Most previous detections of extended coronal gas in low-luminosity radio-quiet AGN resolved CL emission at scales of $<$300~pc \citep{muller+11,rodri+17,may+18}. \citet{ramosa+17} reported that 
the CLR in the teacup galaxy as measured from the narrow component of the line is unresolved ($<$760 pc), but the outflowing component is resolved and its seeing-deconvolved radial size is 860~pc. In this radio-quiet quasar, there seems to be evidence for a jet driving the ionized outflow.
Extended CL gas at kiloparsec scales has already been observed in radio-galaxies.  \citet{tadhunter+88} presented solid evidence of [\ion{Fe}{7}]~$\lambda$6087 emission from a cloud at $\sim$8~kpc from the AGN in PKS\,2152-69. In that source, as well as in other similar radio-loud AGN \citep{stad03}, the effect of induced shocks by a jet/cloud interaction has been claimed as the origin of the extended coronal emission.

The coronal gas kinematics shown in the preceding Section suggest a scenario where the radio-jet inflates ISM gas located along the region where it propagates. Shocks produced by this interaction result in the observed coronal emission. Because of the high-ionization state, the gas is mostly optically thin, allowing us to see both the approaching and receding components of the shells. 

In order to test the above hypothesis, we first explore if photoionization by the AGN is able to explain the large-scale \ion{H}{1} emission detected co-spatially to [\ion{Fe}{7}]. To this purpose, we
estimate the luminosity predicted for the H$\alpha$ line assuming a pure photoionization scenario. The bolometric luminosity of the AGN in Circinus is $10^{10}~$L$\odot$ \citep{oliva+99}. The solid angle covered by the NR and SR region as measured from the MUSE data is, respectively, 0.137 sr and 0.024 sr. Both solid angles summed represent a fraction of 0.01 of all the surface area seen by the  AGN. The integrated luminosity of the H$\beta$ line can be estimated by Equation~\ref{eq1} \citep{osterbrock89}:

\begin{equation}
 L(H\beta) = h\nu _{H_{\beta }}\eta _{e}^{2}fV\alpha _{H_{\beta }}^{eff}
	\label{eq1}
\end{equation}  
 
where $h$ is the Planck constant, $\nu _{H_{\beta }}$ is the line frequency of the $H_{\beta}$, $\eta$ is the electronic density, $V$ is the volume emitted in the region, $f$ is the filling factor and $\alpha _{H_{\beta }}^{eff}$ is the recombination coefficient.

Taking into account a covering factor of 5\% in Circinus, we estimate a H$\beta$ luminosity of  $2.2 \times 10^{39}$~ergs\,s$^{-1}$ (or $5.9 \times 10^{5}~$L$\odot$). Using a theoretical ratio  H$\alpha$/H$\beta$=3.1, it translates to a H$\alpha$ luminosity of $6.9 \times 10^{39}~ergs/s$ ($1.8 \times 10^{6}~$L$\odot$).
From the observed integrated H$\alpha$ flux in the ionization cone co-spatial to the [\ion{Fe}{7}] emission, we derived a luminosity of $5.9\times 10^{39}$~ergs~s$^{-1}$ (1.5 $\times 10^{6}$~L$\odot$). Thus, the AGN is just barely able to power the  H$\alpha$ emission co-spatial to the coronal gas. As  H$\alpha$ extends from the nucleus up to the edge of the datacube,  we conclude that the AGN alone is not sufficient to account for the extended low-ionized emission. This gives further support to the shock-driven coronal gas scenario.

\citet{mingo+12} had already shown that the AGN in Circinus is able to weakly ionize gas up to $\sim$700 pc away, but strong ionization would only be produced within the inner $\sim$200~pc. They showed that the extended emission observed in the [\ion{O}{3}]~$\lambda$5007 line in Circinus, which coincides with that of [\ion{Fe}{7}] (see Figure~\ref{fig:fe7map}, left panel) is most likely caused by a jet-driven outflow, which would be driving shells of strongly shocked gas into the halo of the host galaxy. 

\begin{figure}
    \centering
    \includegraphics[width=\columnwidth]{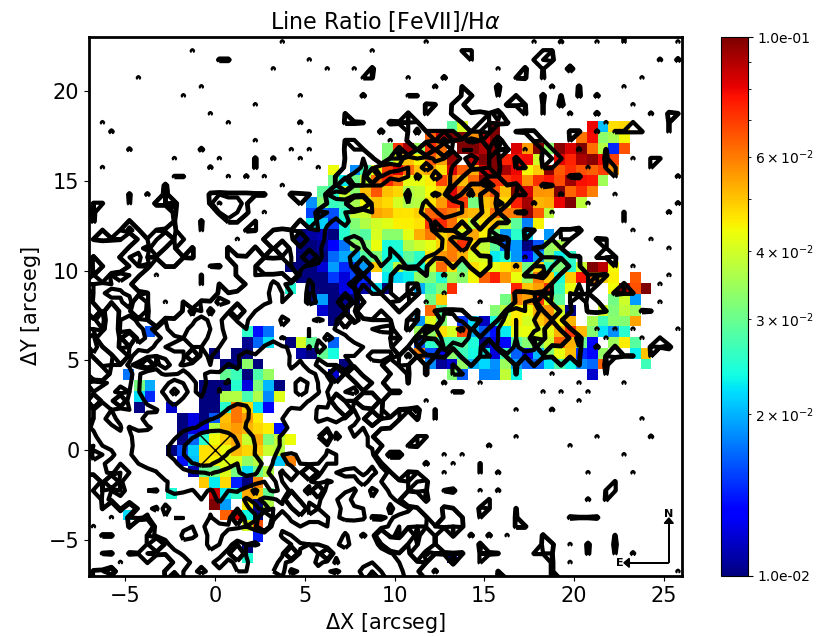}
    \caption{Extinction corrected emission line flux ratio [\ion{Fe}{7}]~$\lambda$6087/H$\alpha$ for the highest ionized portion of the cone. The region in white corresponds to values with S/N $<$ 3 or where the [\ion{Fe}{7}] line is not detected. The cross marks the position of the AGN. The contours represent $Chandra$ ACIS image in the 0.5–8 keV band from \citet{sw01}.}
    \label{fig:fe7o3}
\end{figure}

Additional support to this scenario can be obtained from of the emission line flux ratio [\ion{Fe}{7}]/H$\alpha$ (see Figure~\ref{fig:fe7o3}), which directly reflects the ionization state of the gas. Figure~\ref{fig:fe7o3} shows that the values of that ratio increases outwards, varying from $\sim$0.01 at the side of the NR facing the AGN ($\sim$77~pc) to $\sim$0.1 in the parcel of gas located at 700~pc from the nucleus. A similar effect is also seen in the SR. Here, [\ion{Fe}{7}]/H$\alpha$ increases by a factor of $\sim$6 between the closest and farthest points from the AGN.     

\citet{oliva+99} invoked a large N/O overabundance to explain the rich emission line spectrum in ``knot C" (cloud B, in our Figure~\ref{fig:fe7map}). Although this is indeed a plausible explanation, notice that the [\ion{Fe}{7}]/H$\alpha$ ratio measured in cloud B is nearly a factor of 3 smaller than that in Cloud G. Therefore, such scenario fails at explaining the increasing gas excitation reported here.      
Assuming that the radio-jet is launched from the nucleus, its motion across the ionization cone is highly supersonic, so high-velocity shocks are likely to contribute to the ionization of the line-emitting gas. Long after its passage through the ISM, the gas ionization should decrease due to cooling, while the gas most recently affected by the jet passage should display the largest excitation. The observed distribution of values of the [\ion{Fe}{7}]/H$\alpha$ in Figure~\ref{fig:fe7o3} is consistent with this scenario, as the gas excitation is the largest in the regions most distant from the AGN. 

Theoretical models carried out by \citet{murk+18} and applied to IC~5063, a low-luminosity Seyfert~2 galaxy, shows a highly perturbed ISM by an expanding radio-jet. As the jet floods through the inter-cloud channels of the disc, it ablates, accelerate, and disperse clouds from the central regions of the galaxy, in a similar manner as seen in Circinus.
Moreover, models of \citet{cv01} show that shocks with velocities of 200~km\,s$^{-1}$ and pre-shock densities of $n_{\rm H}$=200~cm$^{-3}$ produce [\ion{Fe}{7}]/H$\alpha$ ratios $>0.1$. Both the FWHM of the lines (see Figure~\ref{fig:kinematics}) and the gas density in the region emitting [\ion{Fe}{7}], of a few hundred cm$^{-3}$ (see \citealt{mingozzi+19}) are consistent with this scenario. 

Finally, the excellent agreement between the extended X-ray emission from $Chandra$ ACIS image in the 0.5–8 keV band \citep{sw01} with the bulk of the [\ion{Fe}{7}] emission, shown in Figure~\ref{fig:fe7o3}, illustrates the strong interplay between AGN jet-driven shocks and the high-ionization lines, not seen before in an AGN. In effect, \citet{mingo+12} interpreted this extended X-ray emission most likely caused by a jet-driven outflow, which is driving shells of strongly shocked gas into the halo of the host galaxy.

This work shows convincing evidence that the jet in the Circinus galaxy is inducing shocks capable of ionizing gas at large distances and possibly pushing outwards the ionized gas. The jet should drive the best resolved and most extended outflow of high-ionization gas ever observed in a low-luminosity AGN, extending up to 700~pc from the central engine. Our work also highlights the use of optical coronal line emission to gather kinematic information not available in the high-ionization gas observed in X-rays via $Chandra$ ACIS data.

\acknowledgments
M.A.F.F. acknowledges the PhD grant from CAPES. A.R.A acknowledges CNPq for partial support to this project. This research has made use of the services of the ESO Science Archive Facility. We thank the Referee for constructing comments/suggestions to improve this manuscript.

%

\vspace{5mm}
\facilities{MUSE/VLT}






\end{document}